\documentclass[runningheads]{llncs}

\usepackage{amsthm, amsmath}

\usepackage{listings}
\usepackage[toc,page]{appendix}
\usepackage[hyphens]{url}
\usepackage{subfig}
\usepackage{algorithm}
\usepackage{algorithmic}
\usepackage{tikz}
\usepackage{comment}
\usetikzlibrary{shapes, arrows}

\tikzstyle{feature} = [circle, text centered, draw=black]
\tikzstyle{table1} = [rectangle, minimum width=3cm, minimum height=0.6cm, draw]
\tikzstyle{table2} = [rectangle split, rectangle split parts=2, minimum height=1.5cm, minimum width=3cm, draw, text centered]
\tikzstyle{table3} = [rectangle split, rectangle split parts=3, minimum height=2.25cm, minimum width=3cm, draw, text centered]
\tikzstyle{table4} = [rectangle split, rectangle split parts=4, minimum height=2.25cm, minimum width=3cm, draw, text centered]
\tikzstyle{table5} = [rectangle split, rectangle split parts=5, minimum height=2.25cm, minimum width=3cm, draw, text centered]
\tikzstyle{table7} = [rectangle split, rectangle split parts=7, minimum height=2.25cm, minimum width=3cm, draw, text centered]
\tikzstyle{table9} = [rectangle split, rectangle split parts=9, minimum height=2.25cm, minimum width=3cm, draw, text centered]
\tikzstyle{arrow} = [thick,->,>=stealth]

\begin{document}

\title{Enriching a CP-Net by Asymmetric Merging}

\author{Stijn Henckens\inst{1}\and
Mostafa Mohajeri Parizi\inst{1}\and
Giovanni Sileno\inst{1}}

\authorrunning{Stijn Henckens, Mostafa Mohajeri Parizi, and Giovanni Sileno}

\institute{Informatics Institute, University of Amsterdam, the Netherlands \email{stijn.henckens@student.uva.nl, \{m.mohajeriparizi,g.sileno\}@uva.nl} 
}

\maketitle

\begin{abstract} 

\noindent Conditional \textit{ceteris paribus} preference networks (CP-nets) are commonly used to capture qualitative conditional preferences. In many use cases, when the preferential structure of an agent is incomplete, information from other preferential structures (e.g. that of other users) preferences can be used to fill in the gaps. Earlier works proposed methods to symmetrically merge multiple incomplete CP-nets by means of voting semantics. However, the merged CP-net can contain preference relations that do not fit to a given user's original preference profile. This paper proposes an asymmetric merging (or \textit{enriching}) method to obtain and fill-in preference relations of a user's CP-net from another CP-net in a way that preserves the original preference relations.


\end{abstract}

\section{Introduction}
Whether it is finding the product that you want, discovering a service that you need, or seeking a person that you share interests with or that you might like, all these tasks rely on personal preferences, typically based-on or influenced-by a wide variety of conditions. Managing these conditional, qualitative preferences is an aspect of growing interest in Computer Science and Artificial Intelligence \cite{walsh2007representing,brafman2009preference}---and this problem becomes even more relevant as people delegate more and more of their tasks to computational agents. A few contributions have indeed started elaborating on how to integrate preferences in autonomous agents \cite{Dasgupta2010,Visser2011,Nunes2014,Mohajeri2019}. Ideally we require these agents to abide by the preferential structure of the associated user. Unfortunately, having the user defining their preferences over all variables can be a tedious or even impossible task.

Also, users might simply be unable to specify their preferences over a continuously growing number of variables 
\cite{wang2016preference}. 
These issues lead to having an incomplete model of preferences that can not be accurately used in applications (e.g. recommender systems, traders, negotiators, or other types of software agents). 

In principle, the problem of incompleteness can be solved by creating an automated process that fills in the missing preferences with plausible choices. Existing methods are based on merging multiple (incomplete) CP-nets \cite{rossi2004mcp,haret2018preference}: the merged CP-net is obtained by aggregation through a voting mechanism in which the input preference structures are weighted equally. However, there exist application domains in which this approach is not the most sound, as for instance when a given preferential structure is preferred over (or deemed ``stronger'' than) the others. This scenario is common in systems of norms. Whereas norms in themselves can be seen as expressing preferential structures (see e.g. \cite{Prakken1997}), systems of norms exhibit internal hierarchies following principles as e.g. \textit{lex posterior derogat priori} (newer norms supersede previous ones) or \textit{lex superior derogat inferiori} (normative sources superior in the institutional hierarchy supersede inferior ones); and plausibly implicit hierarchies due to their relevance with respect to certain underlying values (see e.g. \cite{Bench-Capon2003a}). This difference in requirements motivates an inquiry towards alternative methods: rather than ``aggregating'', the objective targeted here is ``enriching'' an original preferential structure with other preferences, a choice that may bring different (computational) properties and different perspectives on the problem.

More concretely, this work proposes an \textit{asymmetric merge} method to \textit{enrich} a preferential structure specified as a CP-net by using another reference CP-net, that is, to keep all the original (but potentially incomplete) preferences specified in the initial CP-net while inserting all of the preference from the reference CP-net that do not break the consistency of the initial structure.

The paper is organized as follows. In the remainder of this section, we will further analyze the general application requirements in which our contribution can be referred to, and provide references to related works. Section 2 presents the theoretical framework on which our application builds upon. Section 3 introduces in detail the asymmetric merging method, whereas section 4 provides proofs of 
correctness. 
A note on future developments ends the paper.

\subsection{Application Requirements}
\label{sec:ill}

Works on preferences typically abstract the type of preferences they refer to. Modern recommender systems for instance rely mostly on \textit{revealed preferences}, i.e. those that are \textit{ascribed} to users according to observed behaviour. 
In contrast, \textit{motivating preferences}, as those expressed by norms or also by verbalized desires, are deemed to be deliberately \textit{adopted} by the agent \cite[Ch.~6]{Boer2009}. 
Furthermore, motivating preferences are typically provided at higher-level of abstraction (e.g. in terms of outcomes), therefore might offer guidelines of behaviour also in situations that never occurred before. 

A prototypical use case of the asymmetric merging method investigated here is for interactions between motivating preferences (what the agent is committed to) and revealed preferences (what the agent observes from others), where the first are deemed prioritary on the second ones. 
The method is however also useful to capture interactions between revealed preferences, or between motivating preferences, when the \textit{sources} of preferences have a different strength for the agent. Here we give a few illustrative examples of application:
\begin{itemize}
    \item \textit{recommender systems}: a user likes a music broadcaster, so she wants to clone the broadcaster preferential structure as long as it does not override her own;
    \item \textit{social adaptation} (with potential frictions): the agent wants to align with the societal or its group's preferential structure, as long as this is not detrimental according to its own preferences;
    \item \textit{mimetic behaviour} (frictionless social adaptation): the agent adapts to the revealed preferential structure of the others, even if this overrides its own current preferential structure.
\end{itemize}

\subsection{Related Works}

Previous works have investigated methods to merge multiple (incomplete) CP-nets into a single CP-net \cite{rossi2004mcp,grandi2014aggregating,haret2018preference}, based on voting semantics.
This type of merging can be referred to as \emph{symmetric merging}, as every CP-net is treated as equally important, and it results in a CP-net containing a form of averaged conditional preferences. The conditional preference relations found in the merged CP-net can not be used whenever a user has no preferences specified over a given feature, as using these preferences might conflict with the user's personal preference profile, considering the CP-nets are arbitrarily chosen. Furthermore, using voting semantics disables the merging of two CP-nets, since most voting semantics become insignificant when using only two voters. 

Previous work has shown that user preferences are unlikely to be static \cite{pu2003user,chomicki2005monotonic}. Over time, or as new features or values are added to the preference network, a user's preference can get refined or change. Preference revision can be used to perform these changes as new information enters a CP-net. Two types of preference revision are generally distinguished. These are separated in \emph{monotonic} and \emph{non-monotonic} preference revision, both based on classical belief revision semantics \cite{gardenfors1995belief,chomicki2005monotonic}. Belief revision describes the process of how an agent's belief can change when new information is taken into account \cite{sep-logic-belief-revision}. The AGM model, named after its authors, forms the core of belief revision theory \cite{alchourron1985logic}. In the AGM model, beliefs are represented by formal language sentences. The entire belief of an agent is represented by the set of these formal sentences \cite{sep-logic-belief-revision}. In preference revision, these formal language sentences are represented by the conditional preference relations expressed in CP-nets.

\section{Theoretical Framework}

\subsection{Preference Orderings}
A preference, as considered in this paper, is the notion of a \emph{subjective comparative evaluation} in the form of ``User A prefers X to Y''. Here, the \emph{evaluation} concerns practical reasoning on matters of value.
Furthermore, the \emph{comparative} evaluation refers to expressing this evaluation on an item X relative to another item Y \cite{sep-preferences}. Expressing these comparative evaluations is based on two fundamental concepts; \emph{strict preference} ($\succ$) and \emph{indifference} ($\sim$) \cite{hallden1957logic}. A strict preference describes a notion of an item being `better'  than another item, whereas an indifference corresponds to the notion of items being equal in value and therefore equally preferred. These concepts are also the two fundamentals in a preference ordering as considered in this paper.

A preference ordering happens when a user defines his or her comparative evaluations over a set of values. Values of the same sort are collected in a set, represented as a \emph{feature}.
If a preference ordering over values in a feature is defined, a user has sorted values found in that feature, separated by either a strict preference or an indifference. Assume there exists a feature $F$ consisting of the two values $f_1$ and $f_2$. 
A preference ordering over these values can either be $f_1 \succ f_2$ (meaning that $f_1$ is strictly preferred to $f_2$), $f_2 \succ f_1$ ($f_2$ strictly preferred to $f_1$) or $f_1 \sim f_2$ (both values $f_1$ and $f_2$ equally preferred). Strict preferences are considered strong preferences which are irreflexive and transitive. Indifferent preferences are considered weak preferences which are reflexive and negatively transitive.

To complete the notion mentioned in the beginning of the section, a preference is considered as a \emph{subjective} comparative evaluation. Having a user define a preference makes it personally bound to that user, distinguishing it from the ``objective'' sense that the item considered in the preference is actually better than another item \cite{sep-preferences}. Using this notion of a preference, a preference ordering can be seen as a concatenation of subjective comparative evaluations.

Since constructing a preference order relies heavily on a user's personal evaluations and comparisons, it is imaginable that there also exist personally bound factors that influence the comparative evaluation towards a preference. These factors are referred to as \emph{conditions}. It is possible for a preference ordering rely entirely on a given condition. 

\subsection{CP-Nets}\label{CPnet}

The \textit{conditional preference network}  \cite{boutilier1999reasoning,boutilier2004cp} (in short CP-net) is a compact graphical model that can specify the qualitative and conditional relations of preference orderings under \emph{ceteris paribus} semantics. This section will introduce its core definitions.

A CP-net $N$ consists of a set of preferences over a finite set of features $\mathcal{F} = \{F_1, \ldots ,F_n\}$. (Abusing the notation, in the following we will say that $N$ contains a feature $F$ ($F \in N$) to denote that $F \in \mathcal{F}$.)
For every feature $F_i \in \mathcal{F}$ there exists a domain of values, noted by $D(F_i)$. These values are partially ordered and denoted by the operators $\succ$ and $\sim$, which corresponds to ``more preferred'' and ``indifferent'' respectively. The complete ordering of all values in all domains can be expressed in a finite set of possibilities. The set of possible orderings over $\mathcal{F}$ is defined as $\mathcal{O}_\mathcal{F} = \prod_{F_i \in \mathcal{F}} D(F_i)$ \cite{wicker2007interest}.

Since a CP-net holds conditional preferences, two notions are defined. The first notion is that of \textit{preferential independence}. Assume there exist two subset features $X \subseteq \mathcal{F}$ and its complement $Y = \mathcal{F} \setminus X$.

Let $x_1, x_2 \in \mathcal{O}_X$ and $y_1, y_2 \in \mathcal{O}_Y$ be partial outcomes of $X$ and $Y$ and let $x_1y_1$ be the form in which a outcome $x_1y_1 \in \mathcal{O}_\mathcal{F}$ is represented. $X$ is preferentially independent if, and only if, for all $x_1, x_2, y_1, y_2$ it holds that
\begin{align*}
    x_1y_1 \succ x_2y_1 \Leftrightarrow x_1y_2 \succ x_2y_2.
\end{align*}
The relation above says that the preference relation over the values of feature $X$, given that all other feature values remain equal, is the same no matter what values the other features take. If the relation above holds, $x_1$ is preferred to $x_2$ \emph{ceteris paribus}, from which the preference relation $x_1 \succ x_2$ can be expressed.

The second notion is that of a \emph{conditional preferential independence}. To define this notion a third set of features $Z$ is considered, such that $X, Y$ and $Z$ divide up the set $\mathcal{F}$ (assuming that they are all non-empty). Let $z \in \mathcal{O}_Z$ be the ordered set of $Z$. The feature sets $X$ and $Y$ are conditionally preferentially independent given $z$ if, and only if, for all $x_1, x_2, y_1, y_2$ it holds that
\begin{align*}
    x_1y_1z \succ x_2y_1z \Leftrightarrow x_1y_2z \succ x_2y_2z.
\end{align*}
This relation states that the preferential independence of $X$ and $Y$ only holds when $Z$ is assigned $z$ \cite{wicker2007interest,boutilier2004cp}. If so, $X$ and $Y$ are conditionally preferentially independent. From here the conditional preference relation $z : x_1 \succ x_2$ can be expressed, where the part before the colon (:) represents the condition.

In a CP-net each feature $F_i \in \mathcal{F}$ is graphically represented by a node. Next to every feature node, the (conditional) preference ordering of the feature values are noted in the \emph{conditional preference table} (CPT). If the ordering of feature values in node $F_1$ depend on feature values in node $F_2$, the CPT of feature $F_1$ contains conditional preference orderings. In that case, $F_2$ is also in the parent set of $F_1$ (noted as $F_2 \in Pa(F_1)$). Graphically this means that for the set $Pa(F_1)$ there exists a directed edge going from each feature node in $Pa(F_1)$ to $F_1$ \cite{boutilier2004cp}.

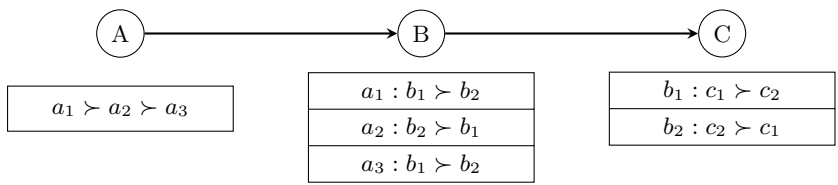
\begin{figure} [h]
\centering
\begin{tikzpicture} [node distance=4cm]
\node (A) [feature] {A};
\node (B) [feature, right of=A] {B};
\node (C) [feature, right of=B] {C};
\node (CPT A) [table1, below of=A, yshift=3cm] {$a_1 \succ a_2 \succ a_3$};
\node (CPT B) [table3, below of=B, yshift=2.75cm] {$a_1:b_1 \succ b_2$ \nodepart{second} $a_2:b_2 \succ b_1$ \nodepart{third}$a_3:b_1 \succ b_2$};
\node (CPT C) [table2, below of=C, yshift=3cm] {$b_1:c_1 \succ c_2$ \nodepart{second} $b_2:c_2 \succ c_1$};

\draw [arrow] (A) -- (B);
\draw [arrow] (B) -- (C);
\end{tikzpicture}
\caption{Example CP-net.} \label{fig:example_cp}
\end{figure}

\begin{example}Consider a CP-net $N$, consisting of feature $A$, with \\$D(A)=\{a_1, a_2, a_3\}$, feature $B$, with $D(B)=\{b_1, b_2\}$ and feature $C$, with $D(C)=\{c_1,c_2\}$. The preference order of $B$ relies on $A$, and the preference order of $C$ relies on $B$. Therefore, $Pa(C)=B$ and $Pa(B)=A$. The example CP-net is visualised in Fig. \ref{fig:example_cp}. \end{example}

\begin{definition} A CP-net $N$ over features $\mathcal{F} = \{F_1,\ldots,F_n\}$ is a directed graph over $\{F_1,\ldots,F_n\}$, whose nodes are annotated with conditional preference tables $CPT(F_i)$ for each $F_i \in \mathcal{F}$. Every $CPT(F_i)$ associates the conditional preference orderings of values $u_i \in U$, where $Pa(F_i) = U$. \cite{boutilier2004cp}\end{definition}

\section{Method}

This section introduces the enrichment method proposed by our contribution. Given an \textit{initial} CP-net and a \textit{reference} CP-net, we enrich the former (the initial structure) by merging preference relations from latter (the reference structure). The operation is performed via an asymmetric merge algorithm. The section provides also multiple examples of the algorithm execution to illustrate its functioning.

\subsection{Asymmetric Merging Algorithm}
The asymmetric merging (or enriching) algorithm takes two CP-nets: $N$, the initial structure to be enriched, and $N'$, the reference structure. 
 
It consists of four main operations:
\begin{itemize}
    \item \textit{unfolding} of both input and reference CP-nets $N$ and $N'$, transforming each independent preference relations into multiple conditional preference relations;
    \item adding all the features $F \in N'$ that $F \notin N$ to $N$ along with all of their respective values;
    \item asymmetrical merging of every conditional preference relation found in the unfolded $N'$, into $N$;
    \item folding the merged conditional preference relations of $N$ back to a compact CP-net.
\end{itemize}
The pseudo code for the enriching algorithm is shown in Algorithm \ref{alg:1} and each operation is explained in detail in the following.

\begin{algorithm}[t]
\caption{Asymmetric Merging (or Enriching)}
\begin{algorithmic}
\label{alg:1}
\REQUIRE CP-net structures $N$, $N'$

\STATE Unfold $N$
\STATE Unfold $N'$
\FOR{feature $F \in N'$}
    \IF{$F \notin N$}
        \STATE Add $F$ to $N$
    \ENDIF
    \FOR{preference relation $p' \in CPT(F_{N'})$}
        \STATE 
        (complete or partial) merge $p'$ with $N$
    \ENDFOR
\ENDFOR
\STATE Fold $N$
\RETURN $N$
\end{algorithmic}
\end{algorithm}

\subsubsection{Unfolding and Folding}
Unfolding an independent preference relation essentially means transforming the implicit \textit{ceteris paribus} information, expanding it to multiple explicitly conditioned preference relations. Assuming a CP-net with set of features $\mathcal{F}$, $X \in \mathcal{F}$, $Z=\mathcal{F} \setminus \{X\}$ and $z \in \mathcal{O}_Z$ where each $z$ associates a value to each feature in $Z$. Unfolding means that each independent preference relation $p_i \in CPT(X)$ is expanded to $|\mathcal{O}_Z|$ conditional relations such that $z \in \mathcal{O}_Z$ is the condition and $p_i$ is the preference relation. Vice versa, multiple unfolded conditional preference relations can be folded back to one independent preference relation. This is possible if the conditions contain all values present in the domains of its parent set.

\begin{example} Consider a CP-net with features $\mathcal{F}=\{Y,X\}$ and $D(X)=\{x_1, x_2\}$. If $CPT(Y)$ contains the independent preference relation $\top : y_1 \succ y_2$, where $\top$ is an empty expression, it can be unfolded into the conditional preference relations $x_1 :y_1\succ y_2$ and $x_2 :y_1\succ y_2$. Subsequently, if $CPT(Y)$ contains the conditional preference relations $x_1 :y_1\succ y_2$ and $x_2 :y_1\succ y_2$, it can be folded in to the independent preference relation $\top : y_1 \succ y_2$. \end{example}

\subsubsection{Adding New Features}
In this step if any given feature is found in the reference CP-net but is not present in the set of features of the initial CP-net, the feature and its values gets inherited to the feature set of the initial CP-net. Here the CPT of the newly added feature remains empty until the asymmetric merging takes place (the asymmetric merging will take care of values of a feature which are present in the reference network but not in the initial one).

\subsubsection{Merge}
The merging of preference relations can be done in two ways; \textit{complete} and \textit{partial}. In the following presentation we will consider two CP-nets, $N$ and $N'$, both containing a feature $F$. 

\paragraph{Complete merge}
Performing a complete merge corresponds to completely inherit a conditional preference relation coming from the reference CP-net. A complete merge of a preference relation $p' \in CPT(F_{N'})$ occurs under either of these two conditions: (1) $p'$ has the condition $z$ but there is no $p \in CPT(F_{N})$ with the same condition $z$ or (2) $p'$ is a preference relation over a set of values $U \subseteq F$ but there is no  $p \in CPT(F_{N})$ that defines preference relations over any of the values $u \in U$. Under these conditions merging results to copying the $p'$ to $CPT(F_{N})$. 
\paragraph{Partial Merge}

A partial merge is done to inherit preferences over feature values from a preference relation $p' \in CPT(F_{N'})$ and insert them in the preference ordering found in preference relation $p \in CPT(F_{N})$ while maintaining previously defined preferences.

Partial merge of a preference relation $p' \in CPT(F_{N'})$ where $z$ is the condition of $p'$ occurs if (1) $p$ can not be completely merged and (2) there is a $p \in CPT(F_{N})$ with condition $z$ that defines a relation containing at least one of the values in $p'$.

\begin{algorithm}[h]
\caption{Partial Merge}
\begin{algorithmic}
\label{alg:2}
\REQUIRE $p' \in CPT(F_{N'})$, $p \in CPT(F_{N})$
\FOR{value $x \in p'$ and $x \notin p$}
    \IF{corresponding position of $x$ in $p$ according to $p'$ is found} 
        \STATE insert $x$ into $p$ in correct position 
    \ENDIF
\ENDFOR
\RETURN $p$
\end{algorithmic}
\end{algorithm}

In Algorithm \ref{alg:2}, corresponding means that after $x$ is added to this position in $p$, for each feature value $f \in p'$, that has $f \in p$, if a preferential relation holds between $f$ and $x$ in $p'$, the same relation should also hold in $p$.

\begin{example}\label{partialmergeexample}
Consider these two conditional preference relations: 
\begin{align*}
    b_1 &: a_1 \succ a_2 \succ a_3\ (p)\\
    b_1 &: a_5 \succ a_3 \succ a_4 \succ a_1 \succ a_2\ (p').
\end{align*}
It is clear that $p$ and $p'$ can not merge entirely, since the preference ordering $a_1 \succ a_2 \succ a_3$ in $p$ is not found in $p'$. However, it is also apparent that $p'$ holds the value $a_5$. This value is preferred over all other values in $p'$, including the values $a_1, a_2$ and $a_3$. Following the partial merging approach, $a_5$ can be inserted into $p$. This means that for every value in both $p'$, that also exists in $p$, its preference relative to $a_5$ is checked. This results in the following preference relation:
\begin{align*}
    b_1 : \{\emptyset\} \succ a_5 \succ \{a_3, a_1, a_2\}
\end{align*}

Now that the position of $a_5$ is found in $p$, $a_5$ can be inserted into $p$ in the same position, resulting in
\begin{align*}
    b_1 : a_5 \succ a_1 \succ a_2 \succ a_3\ (p)
\end{align*}
Continuing, $p'$ also holds another value $a_4$ not present in $p$. Repeating the process described above for value $a_4$ according to $p'$ a position for $a_4$ needs to satisfy: 
\begin{align*}
    b_1 : \{a_5, a_3\} \succ a_4 \succ \{a_1, a_2\}
\end{align*}
But by referring to $p$ it becomes clear that such position does not exist since in $p$ a value can not be strictly preferred to $a_1$ or $a_2$ and less preferred to $a3$ at the same time. Therefore it is concluded that $a_4$ will not be included in the merging process. The final enriched result for $p$ will be:
\begin{align*}
    b_1 &: a_5 \succ a_1 \succ a_2 \succ a_3\ (p).
\end{align*}
\end{example}

\begin{example}
Consider the following conditional preference relations: 
\begin{align*}
    b_1 &:a_1\succ a_2 \succ a_3\ (p)\\
    b_1 &:a_5 \succ a_2 \succ a_3\ (p')
\end{align*}
Following the partial merge approach, $p'$ can be merged with $p$ and the position of $a_5$ in $p$ is determined. The preference relation of the mutual values in $p$ and $p'$ organised relative to $a_5$ is shown below.
\begin{align*}
    b_1:\{\emptyset\}\succ a_5 \succ \{a_2, a_3\}
\end{align*}
This means that $a_5$ gets assigned the same position in $p$ as $a_1$. There is no specified preference between $a_1$ and $a_5$. Therefore, a new indifferent preference relation is created and added to the initial CP-net. The indifferent preference relation is shown below.
\begin{align*}
    b_1:a_1 \sim a_5\ (p)
\end{align*}
\end{example}

\begin{example}
The combination of adding new features, preference relations and/or single values is supported by the asymmetric merging algorithm. Fig.~ \ref{fig:merge} illustrates an example of an entire asymmetric merge between two CP-nets.

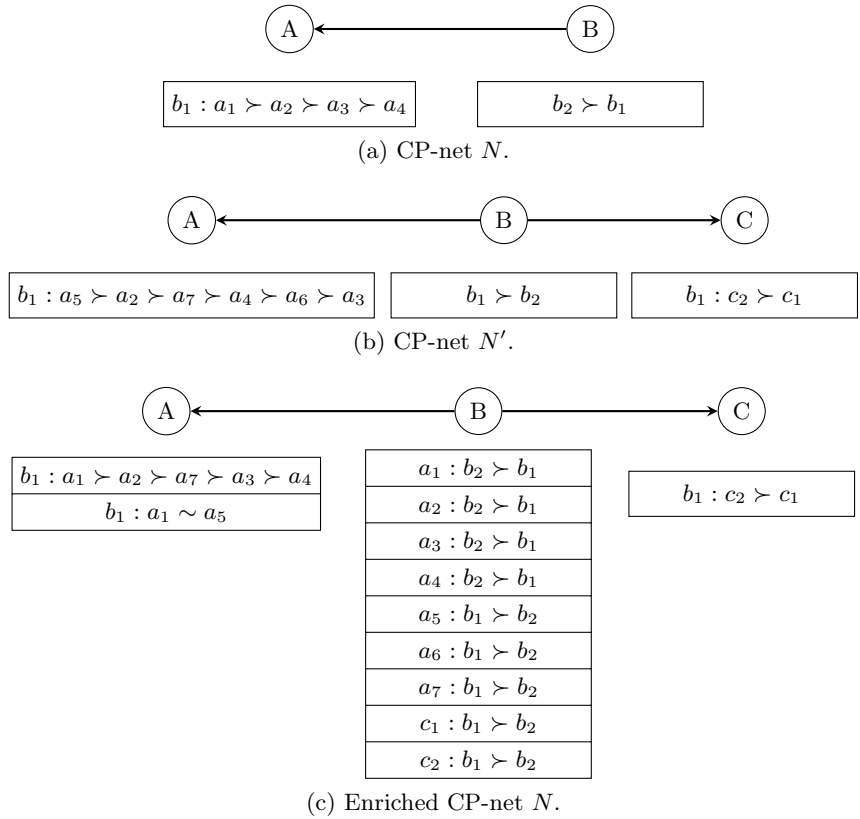
\begin{figure}[h]
\centering
\subfloat[CP-net $N$.] {
\begin{tikzpicture} [node distance=4cm] 
\node (A) [feature] {A};
\node (B) [feature, right of=A] {B};

\node (CPT A) [table1, below of=A, yshift=3cm] {$b_1:a_1 \succ a_2 \succ a_3 \succ a_4$};
\node (CPT B) [table1, below of=B, yshift=3cm] {$b_2 \succ b_1$};

\draw [arrow] (B) -- (A);
\end{tikzpicture}
}\hfill
\subfloat[CP-net $N'$.] {
\begin{tikzpicture} [node distance=4.5cm] 
\node (A) [feature] {A};
\node (B) [feature, right of=A, xshift=-.35cm] {B};
\node (C) [feature, right of=B, xshift=-1.3cm] {C};

\node (CPT A) [table1, below of=A, yshift=3.5cm] {$b_1:a_5 \succ a_2 \succ a_7 \succ a_4 \succ a_6 \succ a_3$};
\node (CPT B) [table1, below of=B, yshift=3.5cm] {$b_1 \succ b_2$};
\node (CPT C) [table1, below of=C, yshift=3.5cm] {$b_1:c_2 \succ c_1$};

\draw [arrow] (B) -- (A);
\draw [arrow] (B) -- (C);

\end{tikzpicture}
} \hfill
\subfloat[Enriched CP-net $N$.] {
\begin{tikzpicture} [node distance=4.5cm] 
\node (A) [feature] {A};
\node (B) [feature, right of=A, xshift=-.35cm] {B};
\node (C) [feature, right of=B, xshift=-1cm] {C};

\node (CPT A) [table2, below of=A, yshift=3.4cm] {$b_1:a_1 \succ a_2 \succ a_7 \succ a_3 \succ a_4$ \nodepart{two}$b_1:a_1 \sim a_5$};

\node (CPT B) [table9, below of=B, yshift=1.8cm] {$a_1: b_2 \succ b_1$ \nodepart{two}$a_2 : b_2 \succ b_1$\nodepart{three}$a_3 : b_2 \succ b_1$\nodepart{four}$a_4 : b_2 \succ b_1$\nodepart{five}$a_5: b_1 \succ b_2$ \nodepart{six}$a_6 : b_1 \succ b_2$\nodepart{seven}$a_7 : b_1 \succ b_2$\nodepart{eight}$c_1 : b_1 \succ b_2$\nodepart{nine}$c_2 : b_1 \succ b_2$};

\node (CPT C) [table1, below of=C, yshift=3.4cm] {$b_1:c_2 \succ c_1$};

\draw [arrow] (B) -- (A);
\draw [arrow] (B) -- (C);

\end{tikzpicture}
}
\caption{\textbf{Three CP-nets.} CP-net $N$ (a) is enriched by merging with CP-net $N'$ (b), resulting in the enriched CP-net $N$ (c).} \label{fig:merge}
\end{figure}
\end{example}

\subsubsection{Worst-case time complexity}
The worst-case time complexity of the asymmetric merging algorithm is determined by going over the algorithm step by step, and computing the worst-case time complexity for each step. 
The algorithm starts with unfolding each CP-net. 
In worst-case, it is assumed that every preference relation in the CP-net's CPT is independent. Therefore, a CP-net with $n$ features containing $n$ independent preference relations, unfolds into $n\times n \times n$ conditional preference relations. Assuming a simplistic linear search, this can be executed in cubic time $\left(O(n^3) \right)$.
The algorithm continues going through all features in the CP-net. All operations within the first loop are executed in linear time $\left(O(n) \right)$. The second loop runs $O(n^2)$ times. In this loop, the operation with the worst-case time complexity is a partial merge. The partial merge requires another loop through $n$ values in a conditional preference relation. If a value is merged, the conditional preference relation is updated and the loop over the value resets. This is executed in quadratic time.
The last steps in the asymmetric merging algorithm is folding the CP-net. Just as unfolding, folding is done in cubic time. The worst-case time complexity of the asymmetric merging algorithm is shown in Equation \ref{timecomplexity}.
\begin{align}\label{timecomplexity}
    T_{enrich} &= T_{unfold} + T_{merge} + T_{fold}\\
    &= O(2n^3) + \left(O(n) \times O(n^2) \times O(n^2) \right) + O(n^3)
    = O(n^5) \nonumber
\end{align}

\section{Proof}
Correctness will be proven against two falsifiable constraints, representing the essential desired properties of the output that we expect from asymmetric merging. The constraints are as follows. When enriching CP-net $N$ with $N'$;
\begin{enumerate}
    \item[(1)] A strict preference order specified in CP-net $N$ is never overwritten by a preference order coming from $N'$. Therefore all strict preference orderings in $N$ exist after enrichment.
    \item[(2)] All non-conflicting preferences found in $N'$ exist in $N$ after enrichment.
\end{enumerate}

These constraints can be verified by going through all merging possibilities. The constraints are considered valid if there does not exist an example that can falsify the constraints.

This paper describes two ways to merge an individual preference relation with a given CP-net: complete merge and partial merge. Both are tested against the two constraints. 

A complete merge copies the entire conditional preference relation from $N'$ to $N$. The restrictions for executing a complete merge confirms both constraints. A complete merge never overwrites or rearranges an existing preference since a complete merge adds an entire non-conflicting preference relation to a CPT $(1)$. Subsequently, a complete merge can only contain non-conflicting preference information $(2)$.

A partial merge is executed as soon as it does not pass the requirements for a complete merge. This means that a partial merge can only happen if the condition of a preference relation in $N'$ already exists in a preference relation of $N$. The partial merge creates a new preference relation for every value in $p'$ that is not in $p$ (see Example \ref{partialmergeexample}). This means that the values that are partially merged with $p$ are never values that already exist in $p$ $(1)$. The newly created preference relation for every value in $p'$ is checked for conflicts with the corresponding preference relation $p$. This means that only if there is no conflict, the value coming from $p'$ is added to $p$ $(2)$. Since this holds for every unique value in $p$, a partial merge can not falsify the constraints.

\section{Discussion}

\paragraph{Comparison with Voting Systems}

As we said in the introduction, the vast majority of contributions in the literature on aggregation of CP-nets are investigations on voting semantics \cite{rossi2004mcp,grandi2014aggregating,haret2018preference}. In principle, the notion of asymmetric merging can be mimicked with voting by considering 3 agents: two following the initial CP-nets (the one to be enriched), and the remaining one following the reference CP-net (the one enriching). However, this approach would require some extension to the classic voting scenario, that usually considers the agents to express their preferences on the same features/values. Second, it would not provide computational advantage as the algorithm will treat the three CP-nets independently, even if two of them are just the same. 

\paragraph{Meta-Preferences}
Throughout the algorithms presented in this paper it is always assumed that there is already a reference preferential structure to enrich the initial structure with. This raises an important question: How do we find, and select, possibly generate
the reference CP-net? This assumption was meant to contain the focus of this work but the importance of \textit{meta-preferences} still remains. Two problems can be identified: \textit{reference} (how to denote a certain preferential structure) and \textit{grounding} (how to have access to its content). The examples of applications in section \ref{sec:ill} offer relevant use cases. In the \textit{social adaptation} scenario, for instance, we need to refer to some kind of \textit{statistical norm} of all the preferences in the society of agents. Aggregation methods as those based on voting semantics could provide methods for the grounding phase. In the \textit{mimetic behaviour} case the agent might choose another specific agent  (seemingly) able to capture higher \textit{utility} in society. Still we need some way to define the terms/dimensions for which a certain agent has more utility, that in turn should intuitively depend on the original preferential structure of the agent.

\paragraph{Monotonic and Non-Monotonic Preference Revision}

Enriching a CP-net can significantly change its structure. This change in structure might suggest changes in (conditional) preferences as well. Some preference changes might lead to different preference conclusions, as for instance in Example 3, in which a higher-order preference ($a_5$) is added on top of the higher-order preference of the initial CP-net ($a_1$).

The AGM model specifies three forms of change in beliefs \cite{sep-logic-belief-revision}: \emph{contraction}, a belief is removed from the complete set of beliefs; \emph{expansion}, a belief is added to the set of beliefs; \emph{revision}: a new belief is added to the set of beliefs, while at the same time earlier beliefs are removed to preserve an overall belief consistency. Similarly, in preference revision, preference relations are revised with other preference relations, yielding new preference relations \cite{chomicki2005monotonic}. Since this is exactly what happens during the proposed enriching method, we analyze it according to these categories.

In complete merging a conditional preference relation is added as it is to the merged CP-net; as this new information is added with no additional operation, this is  

a clear example of expansion (also as amonotonic preference revision). 
However, with the proposed method, non-monotonic preference revision can occur as well. 
This can happen in two occasions. The first and most clear example is: after obtaining a new preference relation, a previously existing indifferent preference relation is removed  (thus a contraction). Secondly, 
revisions can occur when the structure of a CP-net needs to be changed in order to maintain consistency. Adding new features and/or values to a CP-net can then result in either gaining, losing, or gaining and losing feature independence.

\begin{figure} [t]
\centering
\subfloat[CP-net $N$] {
\begin{tikzpicture} [node distance=1.25cm] 
\node (A) [feature] {A};
\node (B) [feature, right of=A] {B};
\node (C) [feature, right of=B] {C};

\draw [arrow] (A) -- (B);
\draw [arrow] (B) -- (C);
\end{tikzpicture}
}\hspace{2cm}
\subfloat[CP-net $N'$] {
\begin{tikzpicture} [node distance=1.25cm] 
\node (A) [feature] {A};
\node (E) [feature, above of=A] {E};
\node (D) [feature, right of=E, xshift=1.25cm] {D};
\node (C) [feature, below of=D] {C};

\draw [arrow] (C) -- (D);
\draw [arrow] (D) -- (E);
\draw [arrow] (E) -- (A);
\end{tikzpicture}
}
\caption{The enriching of CP-net $N$ (a) by asymmetric merging with CP-net $N'$ (b) results in a cyclic CP-net.} \label{fig:discussion}
\end{figure}
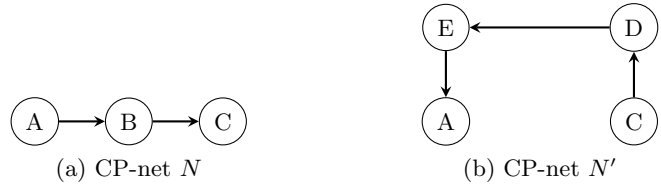

\section{Conclusion and Future Work}

This paper focused on the task of enriching a CP-net with another CP-net by means of asymmetric merging, an alternative (and in many aspect simpler) form of aggregation with respect to those relying e.g. by voting semantics, but that which is suitable to specific niche of applications. Here we suggests two future developments for the proposed method: the introduction of the weaker preference constraints, the introduction of stronger conditional preference relations and a proper treatment of cyclic CP-nets.

\paragraph{Weaker preference constraints}
This paper only considered two types of preferential constraints; a strict preference ($\succ$) and an indifference ($\sim$). The enriching of a CP-net could be extended by considering a weaker preferences ($\succeq$, $\equiv$) as well. These preferential relations translate to "more or equally preferred" and "equally preferred" respectively. 
An example using these additional constraints could be of the following. Consider $p: a_1 \succeq a_2 \succeq a_3$ and $p': a_1 \succeq a_3 \succeq a_2 $. Enriching $p$ with $p'$ would result in $p: a_1 \succeq a_2 \equiv a_3$. Adding these preference constraints in future development could attribute to more detailed enriched CP-nets.

\paragraph{Stronger conditional preferences}
Regardless of the attractive properties a CP-net contains, the \emph{ceteris paribus} semantics houses a downside, as it restricts the use of more stronger preference expressions \cite{wilson2004extending}. In some cases a user wants to define a preference relation where all else is not equal (e.g. "I prefer $f_1$ over $f_2$ regardless of the values of other features"). This from of preference expression proves to be relevant, as it enables a user to define when a feature $F$ is considered the most important feature. On the other hand, it also enables to capture the importance of value preferences by saying that, whenever possible, a given value should be avoided. The implementation of this notion could for example contribute to maintaining desired preference hierarchies, even after enriching.

\paragraph{Cyclic CP-nets}
Adding new features to a CP-net might cause it to become cyclic (e.g. Fig.~\ref{fig:discussion}). 
Cyclic CP-nets can cause problems in consistency and are known to be problematic for several of the most common algorithms available for CP-nets. It would be interesting to research how avoiding cyclic CP-nets could be implemented (e.g. How to decide which feature is the least dependent and could perhaps be excluded from merging?).

\vspace{10pt}
\noindent \textbf{Acknowledgments}
This paper results from work partially done within the NWO-funded project \textit{Data Logistics for Logistics Data} (DL4LD, \url{https://www.dl4ld.net}) in the Commit2Data program (grant no: 628.001.001).

\bibliographystyle{splncs04}
\bibliography{references.bib}

\end{document}